# Investigations of Auditory Filters Based Excitation Patterns for Assessment of Noise Induced Hearing Loss

Wisam Subhi Al-Dayyeni, Pengfei Sun, Jun Qin

Department of Electrical and Computer Engineering,
Southern Illinois University, Carbondale, IL, USA

**Abstract:**

Noise induced hearing loss (NIHL) as one of major avoidable occupational related health issues has been studied for decades. To assess NIHL, the excitation pattern (EP) has been considered as one of mechanisms to estimate movements of basilar membrane (BM) in cochlea. In this study, two auditory filters, dual resonance nonlinear (DRNL) filter and rounded-exponential (ROEX) filter, have been applied to create two EPs, referring as the velocity EP and the loudness EP, respectively. Two noise hazard metrics are also proposed based on the developed EPs to evaluate hazardous levels caused by different types of noise. Moreover, Gaussian noise and pure-tone noise have been simulated to evaluate performances of the developed EPs and noise metrics. The results show that both developed EPs can reflect the responses of BM to different types of noise. For Gaussian noise, there is a frequency shift between the velocity EP and the loudness EP. For pure-tone noise, both EPs can reflect the frequencies of input noise accurately. The results suggest that both EPs can be potentially used for assessment of NIHL.

*Key words*: Noise induced hearing loss; excitation pattern; basilar membrane motion; auditory filter; noise assessment metrics.

## 1. Introduction

Noise-induced hearing loss (NIHL) remains as one of the most common health related problems nowadays as stated by the World Health Organization (WHO). One of the main causes of the permanent hearing loss is the exposure to excessive noise [1-3]. Approximately 22 million workers in the United States are exposed to loud noise workplace that is considered as a hazard level [4]. Hearing loss has a strong impact on the quality of life, causes isolation, impairs social interactions, and increases the risk of accidents [5].

Intrinsically, NIHL can be partially explained as an auditory fatigue phenomenon, in which the motions of stretching and squeezing of basilar membrane (BM) could damage the hearing cells (i.e., outer and inner hair cells) in cochlea [6-8]. The mechanical motions of BM is considered as one of the major factors that causing NIHL in cochlea [9, 10]. The motions of BM in response to the noise stimulus as a function of frequency can be stated as an excitation pattern (EP). Therefore, investigations of the EP are very useful for NIHL research [11].

An EP represents the distribution of movements along BM caused by a sound [12, 13]. In psychoacoustic, the EP is defined as the output of each auditory filter plotted as a function of the filter's center frequency (CF) [14]. The EPs are normally calculated and plotted as the gain of each auditory filter equal to 0 dB at its CF. For example; a tone with a 60 dB sound pressure level (SPL) and at 1 kHz CF will cause an excitation level equal to 60 dB and at 1 kHz [13, 15, 16].

The auditory models (AMs) of the human peripheral frequency selectivity are the fast ways to estimate the EPs over the BM partitions in cochlea [17]. Nowadays many AMs have been developed based on observations of input-output behavior of human auditory system with reference to psychological or physiological responses [6]. Such AMs include Gammatone filters, dual-resonance nonlinear (DRNL) filters, dynamic-compressive gammachirp filters, etc. Hohmann (2002) [18] developed a 4$^{th}$-order linear Gammatone filter based AM for speech processing in hearing aids. This linear model can reconstruct acoustical signals in an auditory system, but it didn't include nonlinear features [18]. Lopez-Poveda and Meddis (2001) [17, 19] proposed a nonlinear DRNL filter, which successfully simulates the two-tone suppression and the phase responses in the BM. Furthermore, Irino and Patterson (2006) [20] developed a gammachirp filterbank with nonlinear and compressive features. The developed gammachirp filter has a group of linear passive gammachirp filters, and can be useful for the applications on speech enhancement, speech coding, and hearing aids [20].

Moreover, the AMs can be categorized as mechanical or perceptual model [21]. The mechanical AMs are designed to estimate mechanical vibrations on BM in cochlea [17], while the perceptual AMs are developed to mimic the psychoacoustic data [20]. In this study, a DRNL filter as a typical mechanical AM and a rounded-exponential (ROEX) filter as a typical perceptual AM have been implemented to investigate EPs on the human BM. As a cascade filter model, the DRNL filter was developed to simulate the nonlinear mechanical response of BM in reaction to stapes motion [19]. The output of DRNL filters is the velocity of BM, which can be described as a velocity EP of BM in cochlea. Such velocity EP intuitively can be used to assess the auditory fatigue based NIHL [6]. On the other hand, the ROEX filter as a perceptual model can be used to describe the loudness levels in cochlea. Loudness is one of the most important parameters for evaluation of the acoustical quality in various applications, from hearing aid optimizing to automatic music mixing systems [22]. The loudness estimations directly reflect the characteristics of human auditory system, such as masking adaption, integration along a perceptual frequency axis, and integration and compression along time axis. In previous studies, loudness contours based models have been developed for evaluations of the annoyance of environment noises, including community noise, industrial noise, and transportation noises [23-25].

In this study, we implement the DRNL filter and the ROEX filter to create two different EPs, the velocity EP and the loudness EP, respectively. To evaluate the performances of both EPs, Gaussian noise and pure-tone noise signals with various parameters (e.g., amplitude and frequency) are simulated. In addition, two noise metrics are proposed based on the velocity EP and the loudness EP to estimate the hazardous levels caused by different types of noise. The rest of this paper is organized as follows. Section 2 describes the auditory filters, the proposed noise metrics, and simulations of noise signals. Section 3 gives experimental results and discussions. Section 4 concludes the paper and outlines future works.

**2. Material and Methods**

2.1. External Ear and Middle Ear

The structure and the function of the auditory system of the human have similarities with other mammalian species. The human ear consists of three parts: external ear, middle ear, and inner ear. Each part in an auditory system plays a unique role to translate acoustic signals from environment to inner ear. The sound passes through external ear in a form of pressure vibration. The pressure vibration changes into mechanical vibration in middle ear. The mechanical energy transforms into hydrodynamic motion in inner ear, and then the BM activates hair cells through electrochemical energy [26].

An external ear consists of ear canal, concha, and pinna flange. The transfer functions of external ear used for the DRNL filter and the ROEX filter are same in this study. As shown in

Fig. 1, the transfer function of external ear is same as it was described in Moore's work [22] and ANSI-S3-2007 [27].

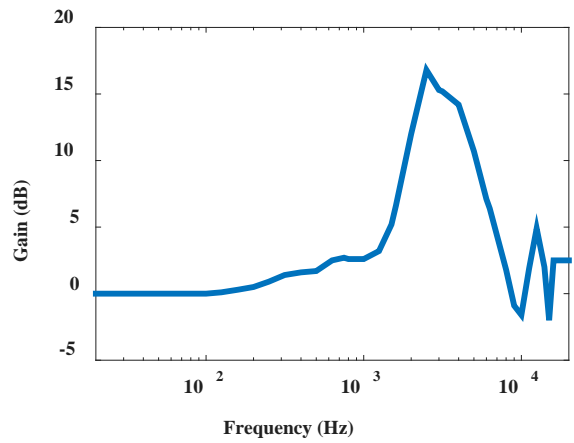

**Fig. 1 - The frequency response of the transfer function of an external ear.**

A middle ear consists of tympanic membrane and ossicular chain, which have three bones. A middle ear plays the role as an impedance-matching device, and it collects and transmits acoustic power to the inner ear [28, 29]. The transfer functions of middle ear have been developed to describe the relationships between inputs and outputs of middle ear [29]. In this study, two different transfer functions of middle ear are applied to the DRNL filter and the ROEX filter, respectively. Fig. 2a shows the frequency response of the transfer function of middle ear used for the DRNL filter as described in Meddis's work [19], in which the acoustical pressure is converted into the stapes velocity, called as the stapes velocity transfer function (SVTF). As shown in Fig. 2b, the frequency response of the transfer function of middle ear for the ROEX filter has been used in the procedure of loudness computation in Moore's work [22].

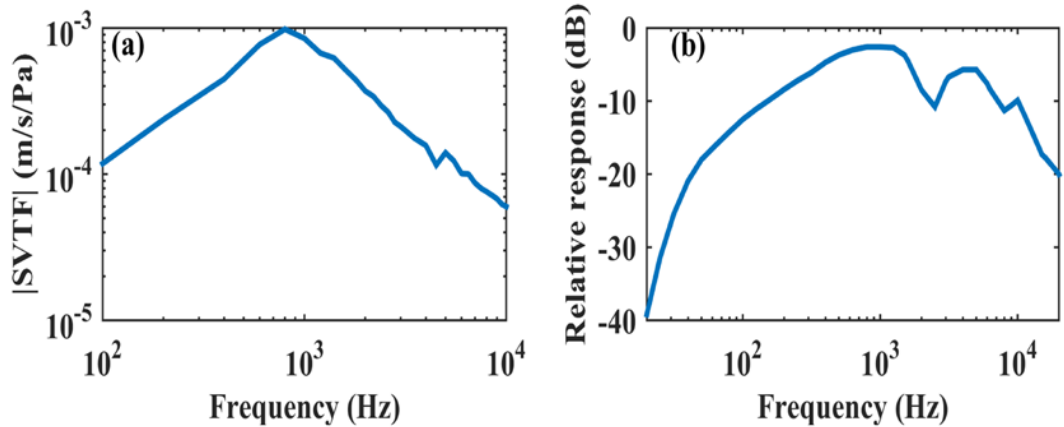

Fig. 2 - The frequency responses of the transfer function of middle ear, which are applied to (a) the DRNL filter [19] and (b) the ROEX filter [22], respectively.

2.2. DRNL filter

In this study, a DRNL filter is utilized to obtain the BM movements in human cochlea [19]. The DRNL filter simulates the velocity of BM as a response to the stapes velocity in middle ear. As shown in Fig. 3, the input of the DRNL filter is the linear stapes velocity. Each individual site is represented as a tuned system with two parallel independent paths, one linear (left) and one nonlinear (right). The linear path consists of a gain /attenuation factor, a bandpass function, and a low pass function in a cascade. The nonlinear path is a cascade combination of the $1^{st}$ bandpass function, a compression function, the $2^{nd}$ bandpass function, and a low pass function. The output of DRNL filter is the sum of the outputs of the linear and nonlinear paths, and is the BM velocity at a particular location along the cochlear partition.

In both paths, each of three bandpass functions consists of a cascade of two or three the $1^{st}$ order gammatone filters [30] with a unit gain at the center frequency (CF). Two low-pass functions are same and consist of a cascade of four $2^{nd}$ order Butterworth low pass filters.

Moreover, the compression function in the nonlinear path was defined based on the animal data, and it can be described as

$$y[t] = \text{SIGN}(x[t]) \times \text{MIN}(a|x[t]|, b|x[t]|)^c \tag{1}$$

where $x[t]$ represents the output from the first bandpass function in the nonlinear path. $y[t]$ is the output of the compression function. $a, b$, and $c$ are models parameters as summarized in Table 1.

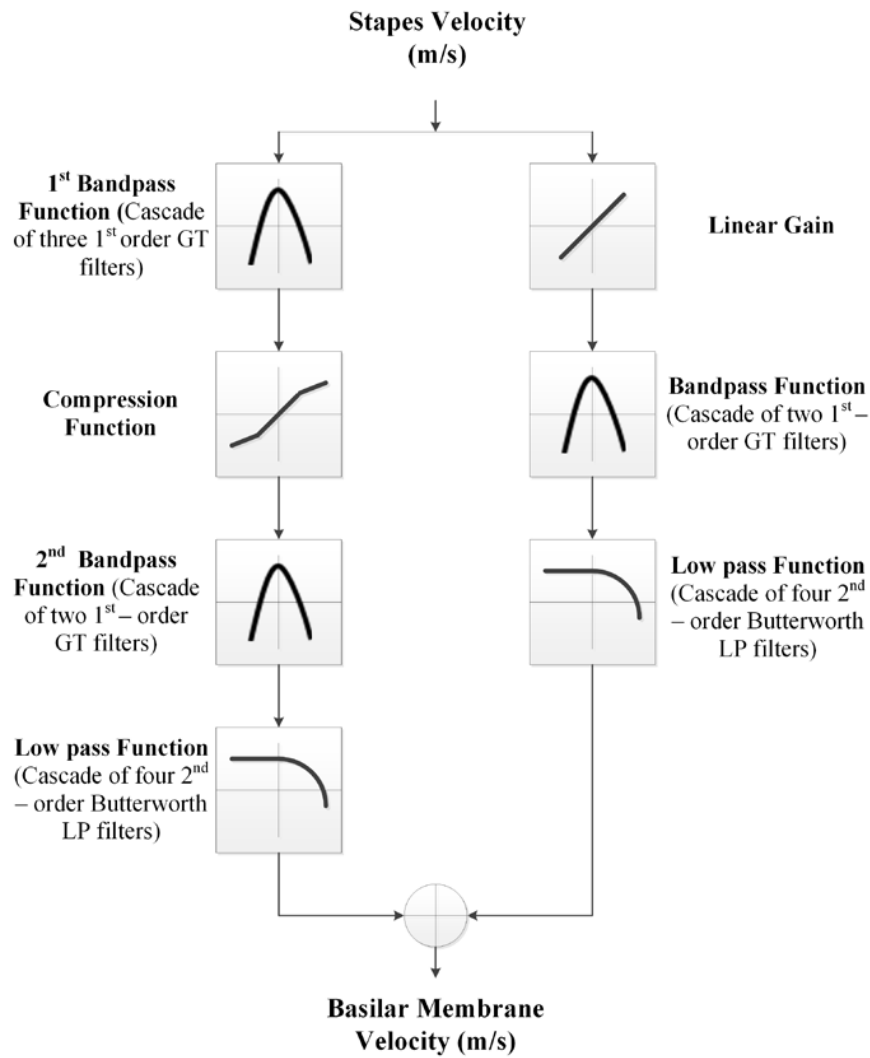

**Fig. 3 - Schematic diagram of the DRNL filter [19], in which the velocities of stapes in middle ear are passed through two parallel branches to obtain the velocities of BM.**

Table 1 summarizes the parameters of the DRNL filter for human used to implement the DRNL filter in this study. The velocity EP is the distribution of BM velocity, which can be obtained as the outputs of the DRNL filter.

**Table 1 - DRNL filter parameters used to simulate the human inner ear [17].**

| Simulated preparation | 0.25kHz | 0.5kHz | 1kHz | 2kHz | 4kHz | 8kHz |
|---|---|---|---|---|---|---|
| Linear | | | | | | |
| GT cascade | 2 | 2 | 2 | 2 | 2 | 2 |
| LP cascade | 4 | 4 | 4 | 4 | 4 | 4 |
| $CF_{lin}$ | 235 | 460 | 945 | 1895 | 3900 | 7450 |
| $BW_{lin}$ | 115 | 150 | 240 | 390 | 620 | 1550 |
| $LP_{lin}$ | 235 | 460 | 945 | 1895 | 3900 | 7450 |
| Gain, g | 1400 | 800 | 520 | 400 | 270 | 250 |
| Nonlinear | | | | | | |
| GT cascade | 3 | 3 | 3 | 3 | 3 | 3 |
| LP cascade | 3 | 3 | 3 | 3 | 3 | 3 |
| $CF_{lin}$ | 250 | 500 | 1000 | 2000 | 4000 | 8000 |
| $BW_{lin}$ | 84 | 103 | 175 | 300 | 560 | 1100 |
| $LP_{nl}$ | 250 | 500 | 1000 | 2000 | 4000 | 8000 |
| Gain, a | 2124 | 4609 | 4598 | 9244 | 30274 | 76354 |
| Gain, b | 0.45 | 0.28 | 0.13 | 0.078 | 0.06 | 0.035 |
| Exponent, c | 0.25 | 0.25 | 0.25 | 0.25 | 0.25 | 0.25 |

2.3. ROEX filter

The ROEX filter was originally derived from psychophysical data [31]. It is a descriptive model, which describes the shape of magnitude transfer function of an auditory filter [32]. The ROEX filter formula can be defined by [27]:

$$W(g) = (1 + pg)exp(-pg) \qquad (2)$$

where $g$ is the normalized deviation from the center frequency (CF) divided by the CF, and $p$ is an adjustable parameter which determines the slope and the bandwidth of the filter.

In this study, the ROEX filter is implemented according to ANSI 3.4-2005 [33]. To calculate the input level at each equivalent rectangular bandwidth (ERB), $p$ in Eq. (3) is set to be $4f/ERB$. The ERB is a psychoacoustic measurement of the width of the auditory filter in each location along the cochlea, and it can be defined as:

$$ERB = 24.673(0.004368f_c + 1) \qquad (3)$$

where $f_c$ is the CF, which are in the range of 50 Hz – 15 kHz in this study. The ERB level obtained according to the input level, which is used to determine the ROEX filter shape. The energy in each ERB can be obtained by:

$$E_i = \frac{\sum W(g_{i,j})P_j^2}{P_0^2} E_0 \qquad (4)$$

where $W$ represents local ROEX filter in the $i$th ERB. $P_j^2$ refers to the power in the $j$th frequency band. $E_0$ is the reference energy at 1 kHz CF and 0 dB SPL, and $P_0$ is the reference pressure referring to $2 \times 10^{-5}$ Pa. For the selected frequencies, $E_i$ will be transformed to loudness levels according to the values of the excitation threshold

$$N = C\left[(G \times E + 2E_{THRQ})^\alpha - (2E_{THRQ})^\alpha\right] \qquad (5)$$

where $E$ is the energy, and G is the low level gain. $C$ and $\alpha$ are two constants, where $C$ =0.046871, and $\alpha$ is related to the $G$ value. $E_{THRQ}$ refers to lower threshold of human perception.

Fig. 4 shows various frequency gains of ROEX filter at different ERB levels. The ROEX filter is a dynamic filter, which has different frequency gains when the levels of ERB change. As the ERB level increases, the slope of the left side of ROEX filter becomes flat. In general, when the sound pressure level (SPL) increases, there will be more energy pass through the ROEX filter. From this perspective, ROEX filter is consistent with the loudness contours. When SPLs increase, loudness contours become flat [14].

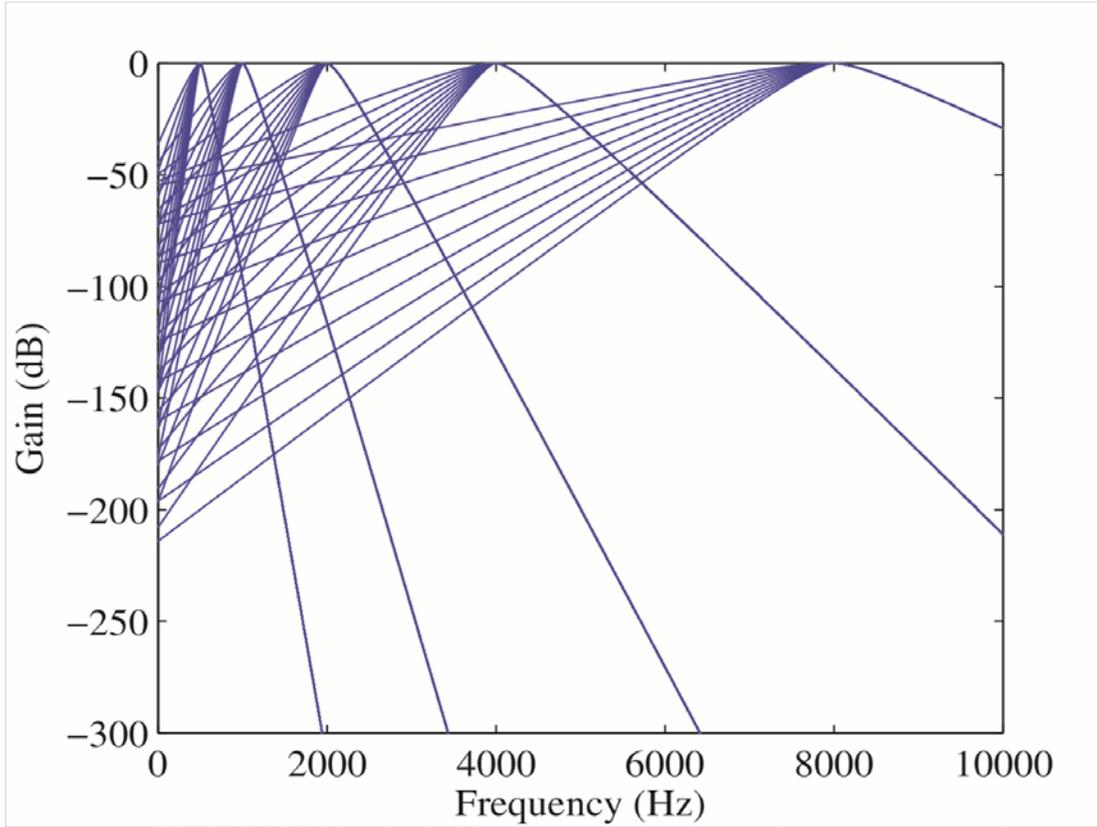

**Fig. 4 - The various frequency gains of ROEX filter centered at 0.5, 1, 2, 4, and 8 kHz at different ERB levels from 20 dB to 100 dB with 10 dB interval.**

2.4. EP Based Noise Metrics

Previous studies have demonstrated that the EPs of BM are highly correlated with NIHL in human cochlea [6, 11, 34]. To investigate hearing loss, two EP based metrics are proposed to assess the potential hazardous levels (HLs) caused by different types of noise. Since the EP represents the temporal responses of the organ of Corti in cochlea, one can integrate the local responses and obtain the cumulative HLs. Therefore, two proposed noise metrics, $HL_i^D$ and $HL_i^R$, can be defined as

$$HL_i^D = 10\log_{10}\sum_{t=1}^{t=n} V(i,t)^2/V_o^2 \tag{6}$$

$$HL_i^R = 10\log_{10}\sum_{t=1}^{t=n} N(i,t)^2/N_o^2 \tag{7}$$

where $HL_i^D$ represents the hazard level index based on the velocity EP, and $V(i,t)$ refers to the BM velocity at the $i$th ERB of BM at time $t$. $V_0$ represents the BM velocity located at the ERB at CF equal to 1 kHz. Moreover, $HL_i^R$ represents the hazard level index based on the loudness EP, and $N(i,t)$ refers to the loudness level at the $i$th ERB of BM at a time $t$. $N_0$ is the loudness level at the ERB at CF equal to 1 kHz. By Eq. (7) and Eq. (8), the developed EPs have been successfully translated to the amount of HLs, which can be potentially used for the assessment of NIHL.

Moreover, total hazard level (THL) can be defined as summation of HLs:

$$THL^D = \sum_i HL_i^D \tag{8}$$

$$THL^R = \sum_i HL_i^R \tag{9}$$

where $THL^D$ and $THL^R$ represent THLs based on the velocity EP and the loudness EP, respectively.

2.5. Simulation of Noise Signals

In this study, two different types of noise signals (i.e., Gaussian noise and pure-tone noise) have been simulated to evaluate the performances of two developed EPs. The Gaussian noise signals are simulated using the "randn" function in MATLAB, in which the probability distribution function of the Gaussian noise is given by [35]:

$$P(t) = \frac{1}{\sigma\sqrt{2\pi}}exp^{-\frac{(t-\mu)^2}{2\sigma^2}} \tag{10}$$

where $\mu$ is the mean, and $\sigma$ is the standard deviation. $\mu$ is equal to zero in this study.

The pure-tone noise signals are simulated by:

$$y(t) = A \cos 2\pi f t \tag{11}$$

where $A$ is the amplitude of the signal, and $f$ is the frequency.

## 3. Results and Discussion

3.1. Time-Frequency (T-F) Representations of Two EPs

In this section, two simulated noise signals (i.e., Gaussian noise and pure-tone noise) are fed into both velocity EP and loudness EP models. The outputs of two EP models are the BM velocity and the loudness level $N(i,t)$ at the $i$th ERB of BM at time $t$, respectively. Both EP can be represented in the joint time and frequency (T-F) domain. Figs. 5a and 5b show the T-F representations of the velocity EP and the loudness EP, responding to a simulated Gaussian noise at 100 dB SPL, respectively. Figs. 5c and 5d show the T-F representations of the velocity EP and the loudness EP, responding to a pure-tone noise with 100 dB SPL and 1 kHz frequency, respectively. The results show that both EPs can reflect amplitudes and transitions of noise signals. The velocity EP as a mechanical model can represent both positive and negative vibrations of BM in cochlea, which reflects more realistic representations of the stretching and squeezing on the hair cells in cochlea. In the other hand, the loudness EP as a perceptual model only represents the positive amount of the loudness as a response to the noise signal. The loudness EP doesn't directly reflect the BM vibrations in cochlea.

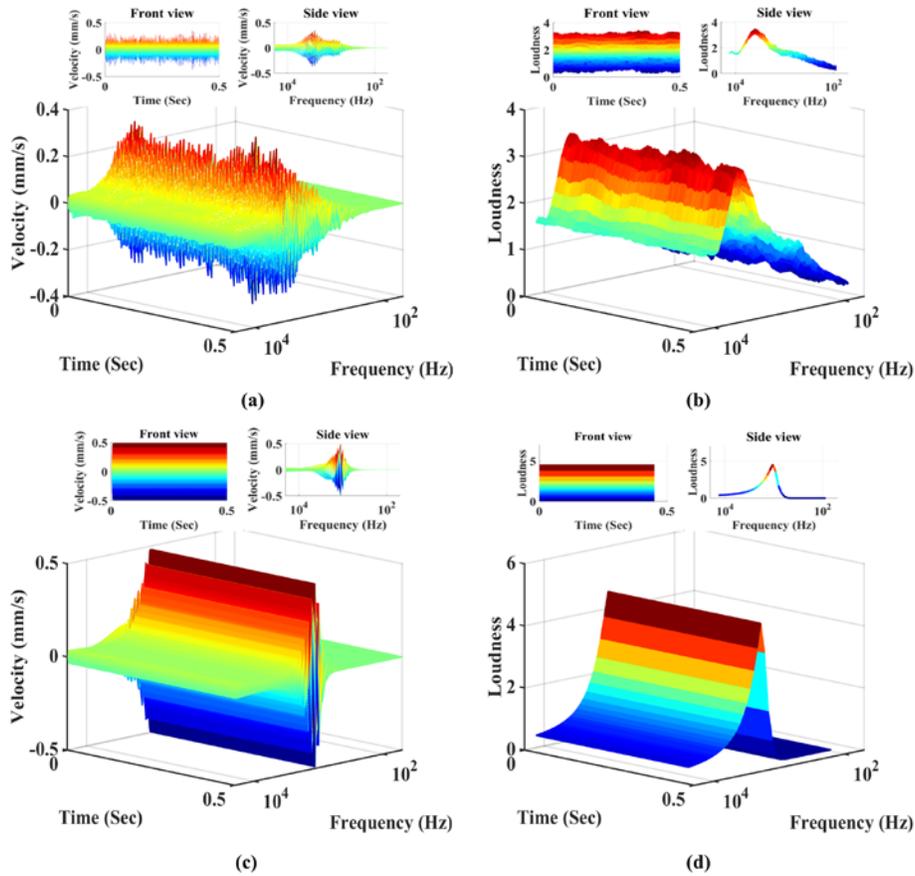

**Fig. 5 - The T-F representations of (a) the velocity EP and (b) the loudness EP responding to a Gaussian noise at 100 dB SPL, and (c) the velocity EP and (d) the loudness EP with respect to a pure-tone noise at 100 dB SPL and 1 kHz.**

Moreover, along the time axis, the velocity EP presents higher temporal resolution than the loudness EP for both Gaussian and pure-tone noise. It indicates that the temporal resolution of the DRNL filter is better than that of the ROEX filter. Along the frequency axis, for the Gaussian noise case, the peak frequency of the velocity EP is around 2 kHz and is lower than the corresponding value of the loudness EP (around 4 kHz). For the pure-tone noise, both EPs present the peak frequencies at 1 kHz, which reflects the frequency of the input pure-tone noise. However, the velocity EP shows vibrations around 1 kHz since it reflects the BM motion while the loudness EP shows only one pulse since it is a perceptual model that reflects the amount of psychoacoustic data.

3.2. T-F Representations of Two EPs for Pure-tone Noise

Fig. 6 shows the T-F representations of the velocity EP and the loudness EP produced by the pure-tone noise signals with 100 dB SPL and various frequencies (i.e., 1, 2, 4, and 6 kHz). For the velocity EP (as shown in the left figures of Fig. 6), the amplitudes have both positive and negative values and the peak amplitudes appear around the frequencies of pure-tone noise signals. It also can be found that the peak amplitudes of the velocity EP are decreasing with the frequency large than 2 kHz. On the other hand, the loudness EP (as shown in the right figures of Fig. 6) presents only positive amplitudes, and the peak amplitudes match the frequencies of stimulating pure-tone noise signals. The peak amplitudes of the loudness EP increase first and then decrease with the frequency increasing, and the maximum peak amplitude appears at 4 kHz.

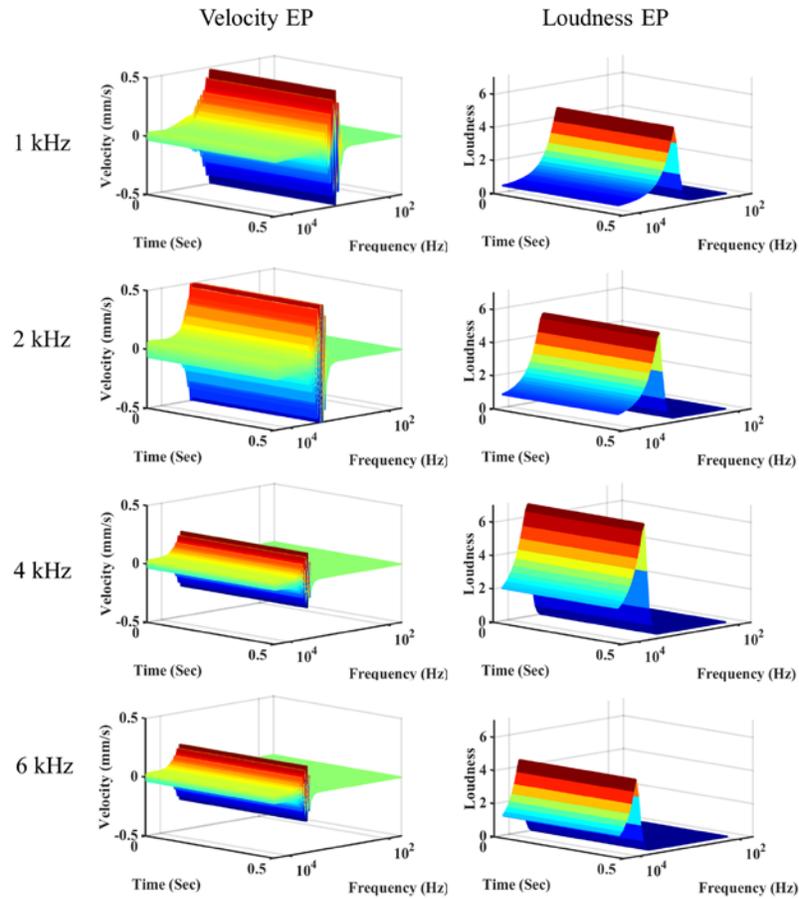

**Fig. 6 – The T-F distributions of two developed EPs obtained by simulated pure-tone noise signals at 100 dB SPL with frequencies at 1, 2, 4, and 6 kHz, respectively.**

3.3. **Hazardous Level Evaluation**

3.3.1. **Frequency Distributions of HLs for Gaussian Noise**

According to Eqs. (6) and (7), the performance two EPs are evaluated using two proposed metrics, $HL_i^D$ and $HL_i^R$, which are used to depict HL at $i^{th}$ ERB on BM. Fig. 7 shows the frequency distributions of normalized HLs generated by the simulated Gaussian noise signals at SPL = 90 to 120 dB with 10 dB interval. For both velocity EP and loudness EP, the HLs increase with SPL increasing. Overall, the loudness EP shows broader frequency response compared with the velocity EP. The results also show that there is a frequency shift between the two EPs. The peak HLs of the velocity EP are around 2 kHz while the peak HLs of the loudness EP are around 4 kHz. Since the BM motions are associated with hearing loss in cochlea, the peak frequency shift between two EPs indicates that the maximum hearing loss predicted by these two EPs may occur at different partitions of BM.

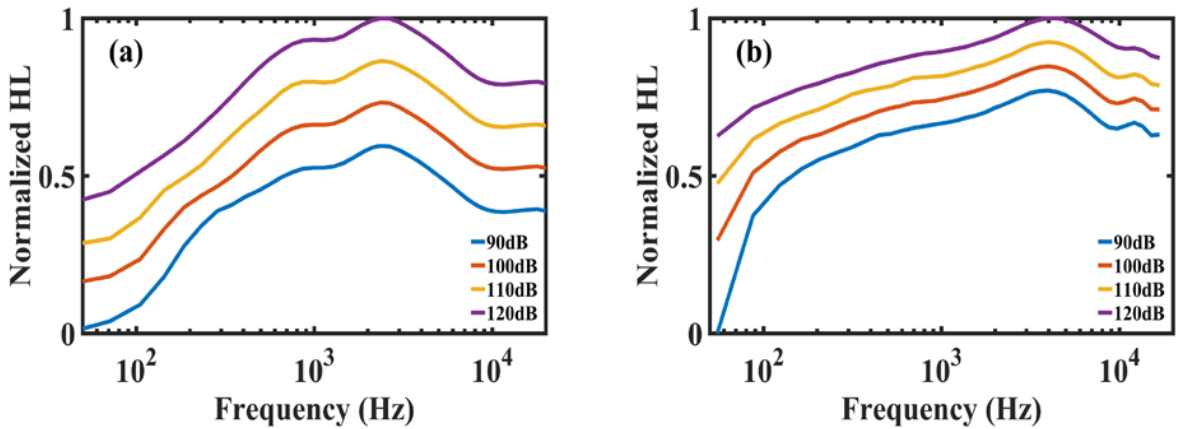

**Fig. 7 - The frequency distributions of normalized HLs based on (a) the velocity EP and (b) the loudness EP generated by simulated Gaussian noise signals at SPL = 90 to 120 dB with 10 dB interval.**

### 3.3.2. Frequency Distributions of HLs for Pure-tone Noise

Fig. 8 shows the normalized HLs generated by simulated pure-tone noise signals at 1 kHz fixed frequency and SPL from 90 to 120 dB with 10 dB interval. Both velocity EP and loudness EP show the peak frequency responses at 1 kHz, which is same as the frequency of the input pure-tone noise signals. It also can be found that the HLs are increasing with SPL levels increasing in both EPs. As shown in Fig. 8a, the HLs of the velocity EP gradually increase when the frequency is smaller than 1 kHz, and then gradually decrease after the frequency is greater than 1 kHz. Comparatively, as shown in Fig. 8b, the HLs of the loudness EP show different frequency responses than the velocity EP. The HLs of the loudness EP almost equal to zero when frequency smaller than 500 Hz, and then rapidly increase with frequency increasing to 1 kHz, and finally gradually decrease with frequency further increasing. This is because the loudness EP is based on the ROEX filter, which is derived from psychophysical data. Therefore, the loudness EP may not reflect the real motion of BM in cochlea.

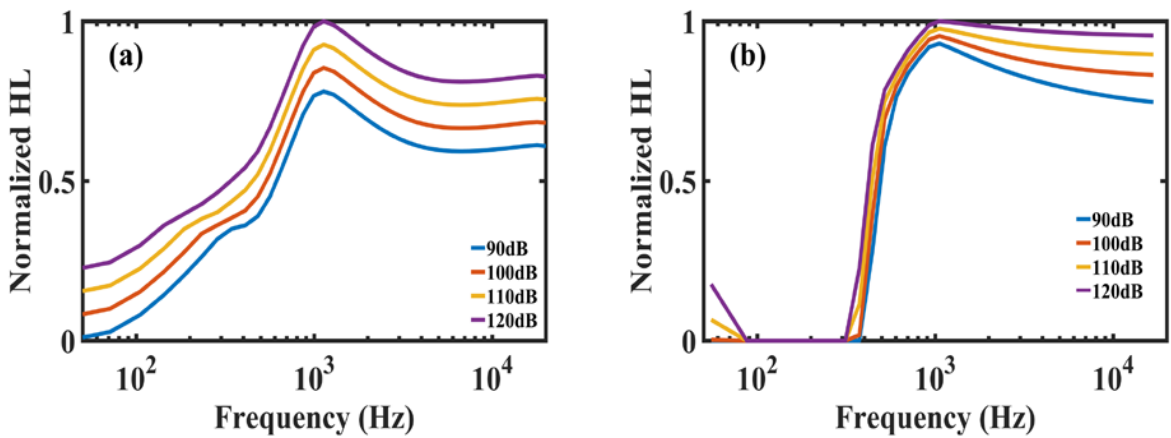

**Fig 8 - The frequency distributions of normalized hazardous levels based on (a) the velocity EP and (b) the loudness EP obtained pure-tone noise at 1 kHz and SPL = 90 to 120 dB with 10 dB interval**.

Moreover, Fig. 9 shows the normalized HLs generated by the simulated pure-tone noise signals at different frequencies (0.5, 1, 2, 4, and 6 kHz) and SPL = 100 dB. Both velocity EP and loudness EP can reflect the corresponding frequencies of input pure-tone noise signals. Specifically, the peak HLs of the velocity EP (as shown in Fig. 9a) is reducing after frequency larger than 2 kHz, while the peak HLs of the loudness EP slightly reduce when frequency higher than 4 kHz. The results in Fig. 9 confirm the peak frequency shift between two EPs generated by

Gaussian noise in Fig. 7. In the velocity EP, the maximum velocity occurs around 2 kHz, while in the loudness EP, the maximum loudness appears around 4 kHz.

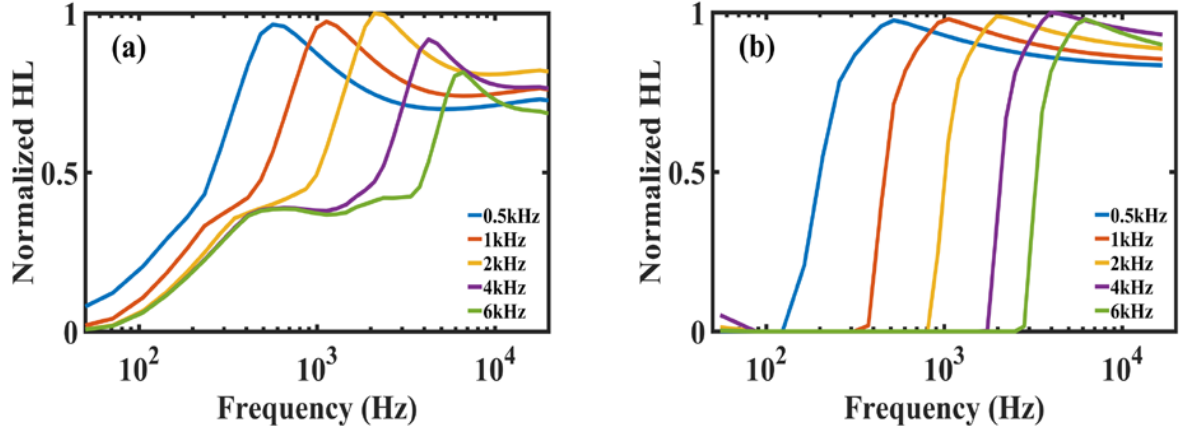

**Fig 9 - The frequency distributions of normalized hazardous levels based on (a) the velocity EP and (b) the loudness EP obtained pure-tone noise signals at various frequencies (0.5, 1, 2, 4, and 6 kHz) with fixed SPL = 100 dB.**

### 3.3.3. Total Hazardous Levels for Gaussian Noise

According to Eq. (8) and Eq. (9), total hazardous levels, $THL^D$ and $THL^R$, can be calculated based on the velocity EP and the loudness EP, respectively. THLs can be used to assess hazardous of high-level noise, and potentially can be used to investigate NIHL. Fig. 10 shows the evaluation of the normalized THLs for the Gaussian noise at SPL from 70 to 120 dB. The result shows that THLs of both EPs are increasing with SPL increasing. The increase rate of the velocity EP is faster than that of the loudness EP. Compared with the loudness EP, the velocity EP shows lower THLs at SPL < 100 dB, but demonstrates higher THLs when SPL > 100 dB.

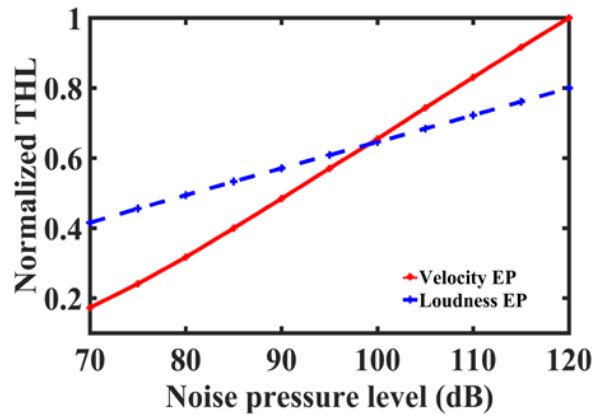

**Fig 10 - The normalized THLs for the Gaussian noise at SPL from 70 to 120 dB for the velocity EP and the loudness EP.**

### 3.3.4. THLs for the Pure-tone Noise

Fig. 11a shows the normalized THL of both EPs produced by the simulated pure-tone noise signals with increasing SPL from 70 to 120 dB and fixed frequency at 1 kHz. The THLs of both EPs are increasing with SPL increasing. Specifically, the velocity EP increases faster than the loudness EP. The result indicates that the velocity EP is more sensitive with SPL increasing than the loudness EP. It also can be found that the THLs of the velocity EP are constantly higher than the corresponding values of the loudness.

Moreover, Fig. 11b shows the normalized THLs of both EPs generated by the simulated pure-tone noise signals at SPL = 100 dB and frequency from 0.5 to 8 kHz. For both EPs, the THLs slightly increase first and then decrease with frequency increasing. The peak THL of the velocity EP is at 2 kHz, while the THL of the loudness EP peaks at 4 kHz. This result is consistent with the previous results in Figs 6, 7 and 9. In addition, the velocity EP shows a fast degradation of THL when the frequency increase more than 2 kHz.

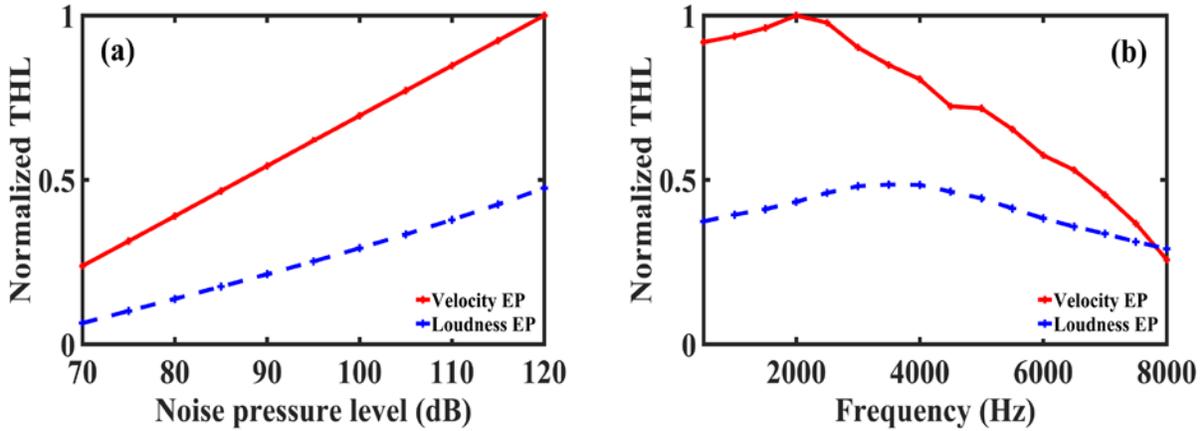

**Fig 11 - The normalized THLs for the pure-tone noise: (a) at 1 kHz and SPL from 70 to 120 dB, and (b) at fixed SPL = 100 dB and frequencies from 0.5 to 8 kHz for the velocity EP and the loudness EP.**

## 4. Conclusions

In this study, two auditory filters, the DRNL filter and the ROEX filter, have been applied to develop the velocity EP and the loudness EP, respectively. Two different types of noise (i.e., Gaussian noise and pure-tone noise) have been simulated to evaluate two developed EPs. For the Gaussian noise, the results show that the maximum velocity obtained by the DRNL filter occurs around 2 kHz, while the peak loudness obtained by the ROEX filter is about 4 kHz. For the pure-tone noise, both EPs can accurately reflect the frequencies of the input noise signals. Moreover, to evaluate the effectiveness of two EPs for prediction of NIHL, we proposed two noise metrics, $HL^D$ and $HL^R$, based on the velocity EP and the loudness EP, respectively. The results show that both EPs can be potentially used as noise hazardous level index for assessment of NIHL. The velocity EP based metric demonstrates higher sensitivity than the loudness EP based metric. However, because the current study is only based on theoretical analysis and simulated noise signals, it may be limited to evaluate the performance of two auditory filters. In our future work, we will utilize experimental animal and human noise exposure data to evaluate the developed velocity EP and loudness EP for assessment of NIHL.